\documentclass[11pt,a4paper,english]{article}
\usepackage{amsmath,amsthm,amssymb}
\usepackage{natbib}
\usepackage{graphicx}
\usepackage{a4wide}
\usepackage{bm}
 \bmdefine\g{g} \bmdefine\K{K}\bmdefine\X{X}
\bmdefine\X{X} \bmdefine\D{D} \bmdefine\K{K} \bmdefine\Z{Z}
\bmdefine\x{x} \bmdefine\z{z} \bmdefine\y{y} \bmdefine\Y{Y}
\bmdefine\bfalpha{\alpha} \bmdefine\bfmu{\mu} \bmdefine\M{M}
\bmdefine\Q{Q} \bmdefine\P{P} \bmdefine\w{w} \bmdefine\W{W}
\bmdefine\p{p} \bmdefine\T{T} \bmdefine\t{t} \bmdefine\r{r}
\bmdefine\B{B} \bmdefine\I{I} \bmdefine\u{u} \bmdefine\p{p}
\bmdefine\Sig{\Sigma} \bmdefine\E{E} \bmdefine\F{F} \bmdefine\S{S}
\bmdefine\s{s} \bmdefine\w{w} \bmdefine\b{b} \bmdefine\W{W}
\bmdefine\w{w} \bmdefine\V{V} \bmdefine\v{v} \bmdefine\q{q}
\bmdefine\R{R} \bmdefine\A{A} \bmdefine\bflambda{\lambda}
\bmdefine\C{C} \bmdefine\U{U} \bmdefine\L{L} \bmdefine\d{d}
\newcommand{\bbeta}{\mbox{\boldmath$\beta$}}

\newcommand{\bSigma}{\mbox{\boldmath$\Sigma$}}

\begin{document}
\title{On the Peaking Phenomenon of the Lasso \\in Model Selection}
\author{Nicole Kr\"amer\\Machine Learning Group, Berlin Institute of Technology\\\texttt{nkraemer@cs.tu-berlin.de}}
\maketitle
\begin{abstract}
I briefly report on some unexpected  results that I obtained when optimizing the model parameters of the Lasso. In simulations with varying observations-to-variables ratio $n/p$, I typically observe a strong peak in the test error curve at the transition point $n/p=1$. This peaking phenomenon is well-documented in scenarios that involve the inversion of the sample covariance matrix, and as I illustrate in this note, it is also the source of the peak for the Lasso. The key problem is the parametrization of the Lasso penalty --  as e.g. in the current \texttt{R} package \texttt{lars} --  and I present a solution in terms of a normalized Lasso parameter.

\end{abstract}
\section{Introduction}
In regression and classification, an omnipresent challenge is the correct prediction in the presence of a huge amount $p$ of variables based on a small number $n$ of observations, and for any regularized method, one typically expects the performance to increase with increasing observations-to-variables ration $n/p$. While this is true in the regions  $n>p$ and $n<p$, some estimators exhibit a peaking behavior for $n=p$, leading to particularly low performance. As documented in the literature \citep{Raudys9801}, this affects all methods that use the (Moore-Penrose) inverse of the sample covariance matrix (see Section \ref{sec:cov} for more details). This leads e.g. to the peculiar effect that for Linear Discriminant Analysis, the performance improves in the $n=p$ case if a set of uninformative variables is added to the model\footnote{Benjamin Blankertz, Ryoto Tomioka: \emph{personal communication}}.
In this note, I show that this peaking phenomenon can also occur in scenarios where the Moore-Penrose inverse is not directly used for computing the model, but in cases where least-squares estimates are used for model selection. One particularly popular method is the Lasso \citep{Tibshirani9601} and its current implementation in  the software $\texttt{R}$. As illustrated in Section \ref{sec:lasso}, its parameterization of the  penalty term in terms of a ration of the $\ell_1$-norm of the Lasso solution and the least-squares solution leads to problems when using cross-validation for model selection. I present a solution in terms of a normalized penalty term.

\section{Simulation Setting and Peaking Phenomenon} \label{sec:lasso}
For a $p$-dimensional linear regression model
\begin{eqnarray*}
y&=& \x^\top \bbeta + \varepsilon\,,
\end{eqnarray*}
the task is to estimate $\bbeta$ based on $n$ observations $\{(\x_1,y_1),\ldots, (\x_n,y_n)\} \subset \mathbb{R}^p \times \mathbb{R}$. As usual, the centered and scaled observations are pooled into $\X= \left(\x_1,\ldots,\x_n\right)^\top \in \mathbb{R} ^{n \times p}$  and $\y= \left(y_1,\ldots,y_n\right)^\top \in \mathbb{R}^n$.

In this note, I study the performance of the Lasso \citep{Tibshirani9601}
\begin{eqnarray*}
\widehat \beta_{\text{lasso}}&=& \text{arg}\min_{\bbeta} \left\{ \|\y- \X \bbeta\|^2 + \lambda \|\bbeta\|_1\right\},\, \lambda \geq 0
\end{eqnarray*}
for a fixed dimensionality $p$ and for a varying number $n$ of observations. Common sense tells us that the test error is approximately a decreasing function of the observations-to-variables ratio $n/p$. However, in several empirical studies, I observe particularly poor results for the Lasso in the transition case $n/p=1$, leading to a prominent peak in the test error curve at $n=p$.

In the remainder of this section, I illustrate this unexpected behavior on a synthetic data set. I would like to stress that the peaking behavior is not due to particular choices in the simulation setup, but only depends on the ratio $n/p$. I generate $n_{\text{total}}=5000$ observations $x_i \in \mathbb{R}^{90}$, where $x_i$ is drawn from a multivariate normal distribution with no collinearity. Out of the $p=90$ true regression coefficients $\bbeta$, a random subset of size $20$ are non-zero and drawn from a univariate distribution on $[-4,+4]$. The error term $\varepsilon$ is normally distributed with variance  such that the signal-to-noise-ratio is equal to $4$. For the simulation, I sub-sample training sets of sizes $n=10,20,\ldots,190,200$. The sub-sampling is repeated $10$ times. On the training set of size $n$, the optimal amount of penalization is chosen via $10$-fold cross-validation. The Lasso solution is then computed on the whole training set of size $n$, and the performance is evaluated by computing the mean squared error on an additional test set of size $500$.

\begin{figure}[htb]
\begin{center}
\includegraphics[width=14cm,height=7cm]{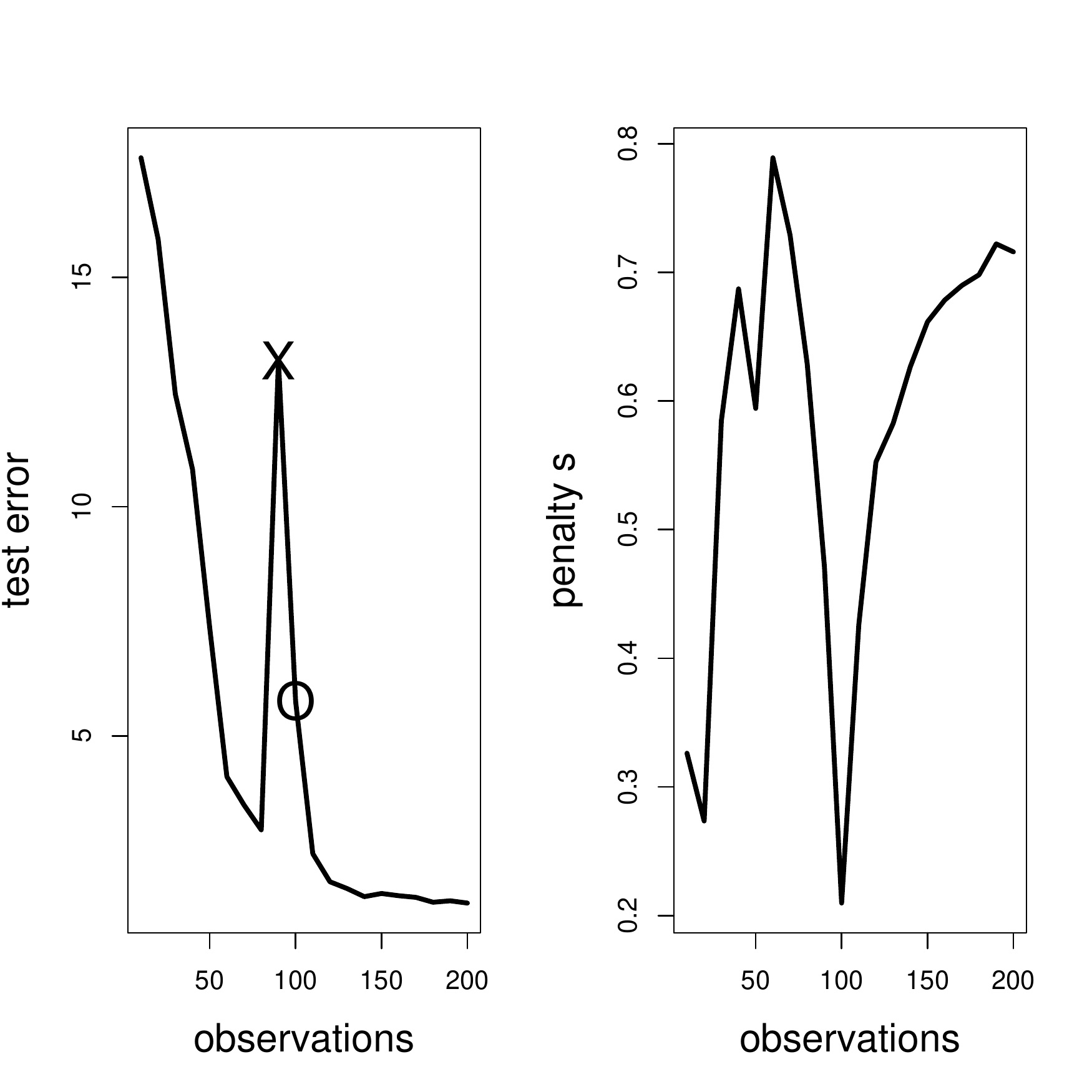}
\end{center}
\caption{Performance of  the Lasso as a function of the number of observations. Left: test error. Right: penalty term $s$ as defined in  Equation (\ref{eq:s}).}

\label{fig:peak_lasso}
\end{figure}

I use the \texttt{cv.lars} function of the \texttt{R} package \texttt{lars} version $0.9-7$ \citep{lars} to perform the experiments. The mean test error over the $10$ runs are displayed in the left panel of Figure \ref{fig:peak_lasso}. As expected, the test error decreases with the number of observations. For $n=p$ however, there is a striking peak in the test error (marked by the letter X), and the performance is much worse compared to the seemingly more complex scenario of $n \ll p$. We also observe the peaking behavior in the case where $n=p$ in the cross-validation split (marked by the letter O). The right panel of Figure \ref{fig:peak_lasso} displays the
cross-validated penalty term of the Lasso as a function of $n$. Note that in the \texttt{cv.lars} function, the amount of penalization is not parameterized by $\lambda \in [0,\infty[$ but by the more convenient quantity
\begin{eqnarray}
\label{eq:s}
s&=& \frac{\|\widehat \beta_{\text{lasso}}\|_1}{\|\widehat \beta_{\text{ols}}\|_1} \in [0,1]\,.
\end{eqnarray}

Values of $s$ close to $0$ correspond to a high value of $\lambda$, and hence to a large amount of penalization. The right panel of Figure \ref{fig:peak_lasso} shows that the peaking behavior also occurs for the amount of penalization, measured by $s$. Interestingly, the peak does not occur for $n=p$, but in the case where the number of observations equals the number of variables in the cross-validation loops. This peculiar behavior is explained in the two following sections, and I also present a normalization procedure that solves this problem.
\section{The Pseudo-Inverse of the Covariance Matrix} \label{sec:cov} %
It has been reported in the literature  \citep{Raudys9801,Tresp,Opper} that the pseudo-inverse of the covariance matrix \begin{eqnarray*}
\widehat \bSigma&=& \frac{1}{n-1} \X^\top \X = \sum_{j=1}^p \widehat \lambda_j \widehat \u_j \widehat \u_j ^\top
\end{eqnarray*}
is a particularly bad estimate for the true precision matrix $\bSigma^{-1}$ in the case $p=n$. The rationale behind this effect is as follows. The Moore-Penrose-Inverse of the empirical covariance matrix is
\begin{eqnarray*}
\widehat \bSigma ^+&=& \sum_{j=1}^{\text{rank}(\bSigma)} \frac{1}{\widehat \lambda_j} \widehat \u_j \widehat \u_j ^\top\,.
\end{eqnarray*}
In particular, in the small sample case, the smallest $p-n$ eigenvalues of the Moore-Penrose inverse are set to $0$.
This corresponds to cutting off directions with high frequency. While this introduces an additional bias, it tends to avoid the huge amount of variance that is due to the inversion of small but non-zero eigenvalues. In the transition case $n/p=1$, all eigenvalues are $\not=0$ (with some of them very small) and the MSE is most prominent in this situation.

The striking peaking behavior for $n=p$ is illustrated in e.g. \cite{Schaefer0501}. As a consequence, any statistical method that uses the pseudo-inverse of the covariance suffers from the peaking phenomenon.

\begin{figure}[htb]
\begin{center}
\includegraphics[width=7cm,height=7cm]{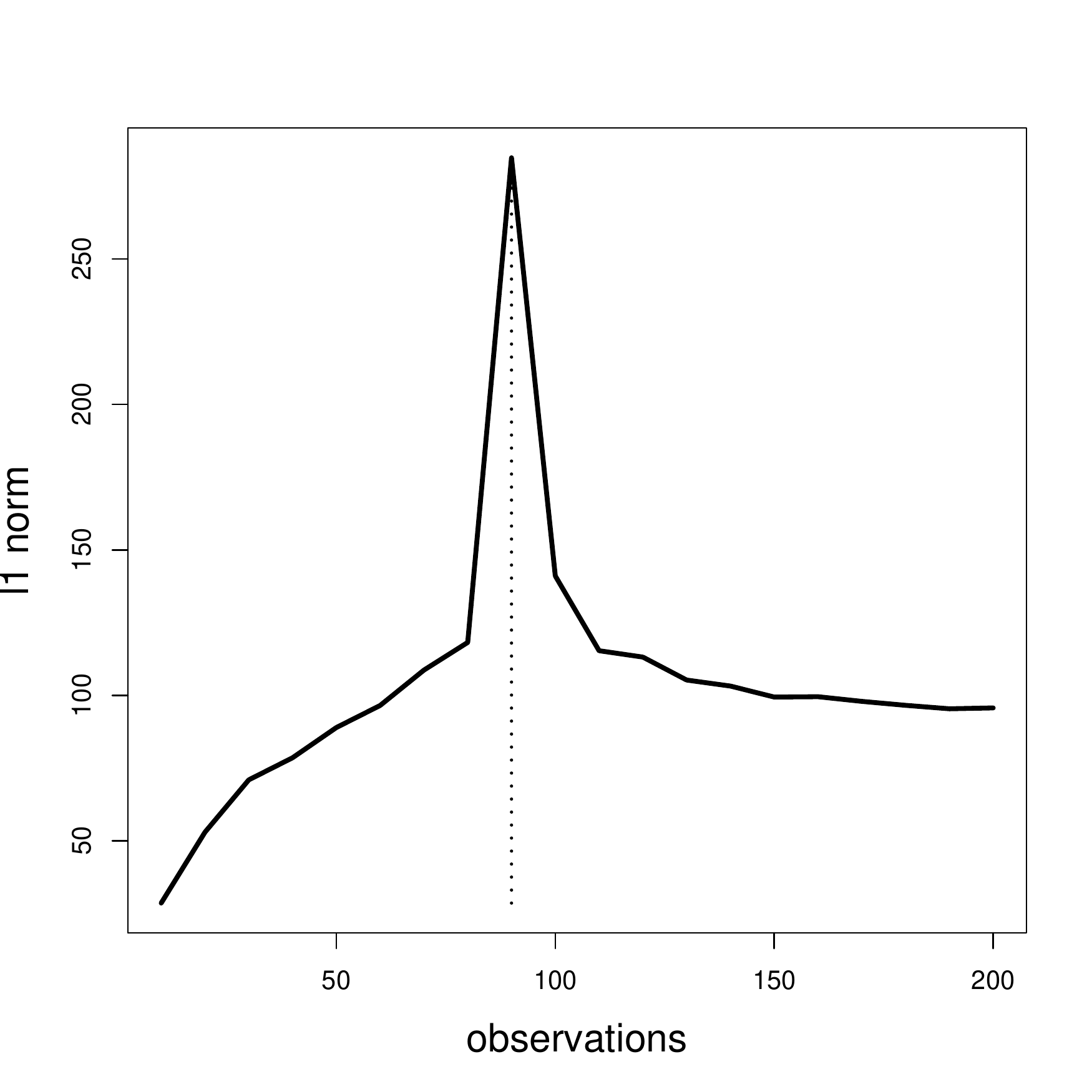}
\end{center}
\caption{Peaking behavior of the ordinary least squares regression: $\ell_1$-norm of the least squares estimate as a function of the number of observations.}

\label{fig:peak_norm}
\end{figure}

consequently, the peaking behavior also occurs in ordinary least squares regression, as it uses the pseudo-inverse,
\begin{eqnarray*}
\widehat \bbeta_{\text{ols}}&=& \left(\X^\top \X\right)^{+} \X ^\top \y\,.
\end{eqnarray*}

 This is illustrated in Figure \ref{fig:peak_norm}. On the training data of size $n=10,20,\ldots,200$, I compute  the $\ell_1$-norm of least squares estimate. The Figure displays the mean norm over all $10$ runs. For $n=p$, the norm is particularly high. Note furthermore that except for $n=p$, the curve is rather smooth, and small changes in the number of observations only lead to small changes in the $\ell_1$-norm of the estimate.

This observation is the key to understanding the peaking behavior of the Lasso. While for the estimation of the Lasso coefficients itself, the pseudo-inverse of the covariance matrix does not occur, it is used for model selection, via the regularization parameter $s$ defined in Equation (\ref{eq:s}). I elaborate on this in the next section.

\section{Normalization of the Lasso Penalty}
Let me denote by $n_{cv}$ the number of observations in the $k$ cross-validation splits, and by $s_{n,cv}$ the optimal parameter  chosen via cross-validation.  As $n \approx n_{cv}$, one  expects the MSE-optimal coefficients $\widehat \bbeta_{\text{lasso},n}$ computed on a set of size $n$ and the MSE-optimal coefficients $\widehat \bbeta_{\text{lasso},n_{cv}}$ based on a set of size $n_{cv}$ to be similar, i.e.

\begin{eqnarray*}
n \approx n_{cv} &\Rightarrow & \|\widehat \bbeta_{\text{lasso},n}\|_1 \approx \|\widehat \bbeta_{\text{lasso},n_{cv}}\|_1\,.
\end{eqnarray*}
Now, if $n_{cv}=p$, then, in each of the $k$ cross-validation splits, the number of observations equals the number of dimensions. As the least squares estimate is prone to the peaking behavior
(recall Figure \ref{fig:peak_norm}), we observe
\begin{eqnarray*}
\|\widehat \bbeta_{\text{ols},n}\|_1 \ll  \| \widehat \bbeta_{\text{ols},n_{cv}}\|_1\,.
\end{eqnarray*}
This implies that even though the $\ell_1$-norms of the regression coefficients $\widehat \bbeta_{\text{lasso}}$are almost the same, their corresponding values of $s$ differ drastically. To put it the other way around: The optimal $s$ found on the cross-validation splits (where $n_{cv}=p$) is way too small, and it dramatically overestimates the amount of penalization. This explains the high test error in the case $n_{cv=p}$ that is indicated by the letter O in Figure \ref{fig:peak_lasso}.

For $n=p$, the same argument applies. The optimal $s_{cv}$ on the cross-validation splits (where $n_{cv} < p$) underestimates the amount of complexity in the $n=p$ case, which leads to the peak indicated by the letter X in Figure \ref{fig:peak_lasso}.

To illustrate that the peaking problem is indeed due to the parametrization (\ref{eq:s}), I normalize the scaling parameter $s$ in the following way. Let me denote by   $\ell_{1,\text{ols}_{cv}}$ the average over all $k$ different  $\ell_1$-norms of the least squares estimates obtained on the $k$ cross-validation splits. Furthermore, $\ell_{1,\text{ols}}$ is the $\ell_1$-norm of the least squares estimates on the complete training data of size $n$. The normalized regularization parameter is
\begin{eqnarray}
\label{eq:snormal}
\widetilde{s}&=& \frac{\ell_{1,\text{ols}_{cv}}}{\ell_{1,\text{ols}}} s_{cv}\,.
\end{eqnarray}
Note that the function \texttt{lars} returns the least squares solution, hence there are no additional computational costs.

\begin{figure}[htb]
\begin{center}
\includegraphics[width=14cm,height=7cm]{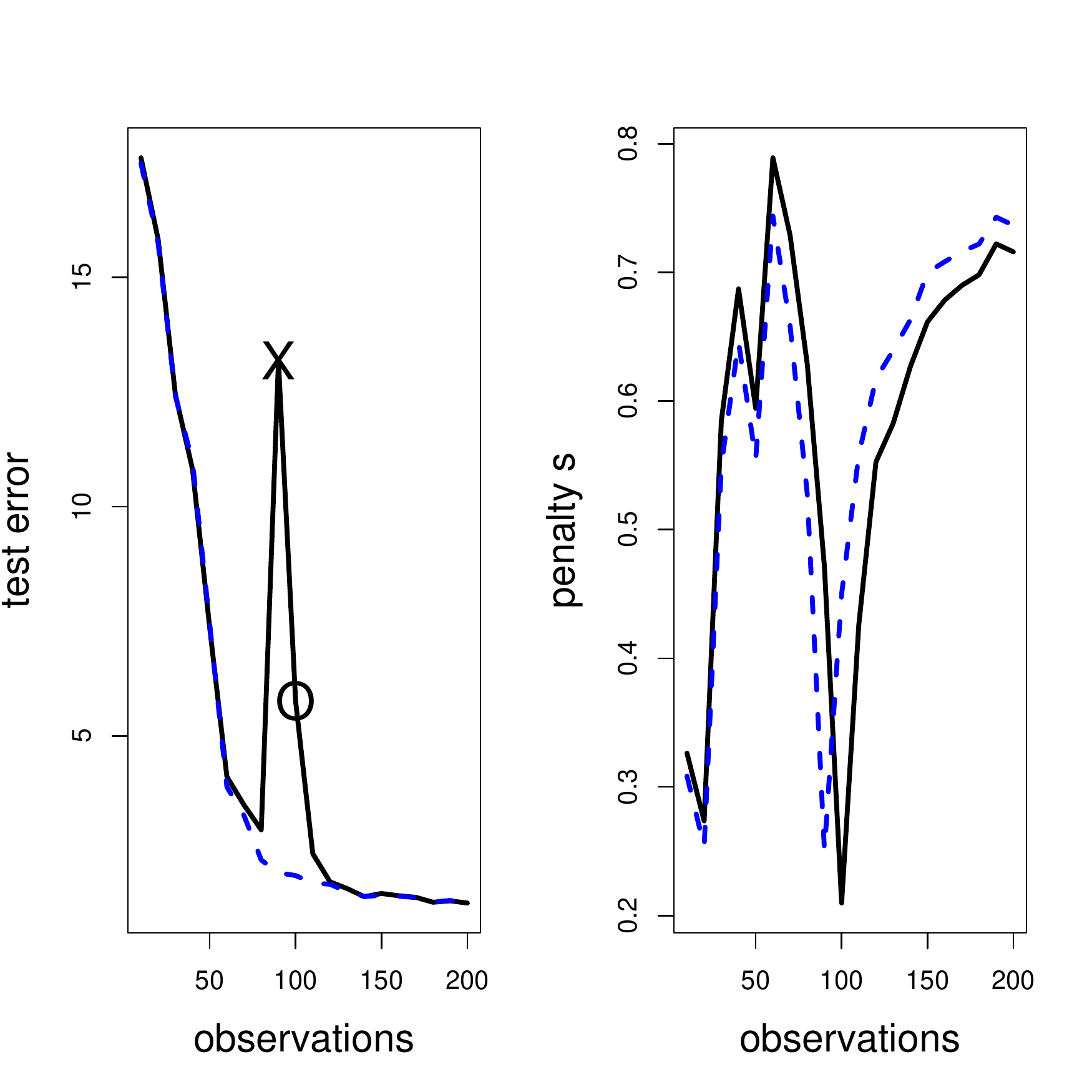}
\end{center}
\caption{Performance of  the Lasso (black solid line) and the normalized Lasso (blue jagged line) as a function of the number of observations. Left: test error. Right: penalty term $s$ (black solid line) and $\widetilde{s}$ (blue jagged line) as defined in  Equation (\ref{eq:s}) and (\ref{eq:snormal}) respectively.}
\label{fig:peak_lasso2}
\end{figure}

To illustrate the effectiveness of the normalization, I re-run the simulation experiments with cross-validation based on the normalized penalty parameter (\ref{eq:snormal}). This function - called \texttt{mylars} -- is implemented in the \texttt{R}-package \texttt{parcor} version 0.1 \citep{parcor}. The results together with the results for the un-normalized parameter \ref{eq:s} are displayed in Figure \ref{fig:peak_lasso2}.

\section{Conclusion}
The  peaking phenomenon  is well-documented in the literature, and it effects every estimator that uses the pseudo-inverse of the sample covariance matrix. As I illustrate in this note, this defect in the transition point $n/p=1$ can also occur in more subtle ways. For the Lasso, the particular parameterization of the penalty term uses least-squares estimates, and it leads to difficulties in model selection. One can expect similar problems if one e.g. measures the fit of a model in terms of the total variance that it explains, and if the total variance is estimated using least squares. In this case, a normalization as proposed above is advisable.

\subsubsection*{Acknowledgements}
I observed the peaking phenomenon during the preparation of a paper with Juliane Sch\"afer and Anne-Laure Boulesteix on regularized estimation of gaussian graphical models \citep{Kraemer0901}. Together with Lukas Meier, the three of us discussed the source of the peaking phenomenon in great detail. My colleagues  Ryota Tomioka,  Gilles Blanchard and  Benjamin Blankertz provided additional material to the discussion and pointed to relevant literature.

\bibliographystyle{apalike}

\end{document}